\documentclass[twocolumn,prb,aps,amsmath,showpacs]{revtex4}

\usepackage{graphicx}
\usepackage[version=3]{mhchem}
\usepackage{dcolumn}
\usepackage{multirow}
\usepackage{bm}
\usepackage{subfigure}
\usepackage{cases}
\usepackage{amsmath}

\begin{document}
\title{Polymorphism and thermodynamic ground state of Silver fulminate studied from van der Waals density functional calculations}
\author{N. Yedukondalu, and G. Vaitheeswaran$^*$ }
\affiliation{Advanced Centre of Research in High Energy Materials (ACRHEM), University of Hyderabad, Prof. C. R. Rao Road, Gachibowli, Hyderabad- 500 046, Andhra pradesh, India }
\date{\today}

\begin{abstract}
Silver fulminate (AgCNO) is a primary explosive, which exists in two polymorphic phases namely orthorhombic (\emph{Cmcm}) and trigonal (\emph{R$\bar{3}$}) forms at ambient conditions. In the present study, we have investigated the effect of pressure and temperature on relative phase stability of the polymorphs using planewave pseudopotential approaches based on Density Functional Theory (DFT). van der Waals interactions play a significant role in predicting the phase stability and they can be effectively captured by semiempirical dispersion correction methods incontrast to standard DFT functionals. Based on our total energy calculations using DFT-D2 method, the \emph{Cmcm} structure is found to be the preferred thermodynamic equilibrium phase under studied pressure and temperature range. Hitherto \emph{Cmcm} and \emph{R$\bar{3}$} phases denoted as $\alpha$ and $\beta$-forms of AgCNO, respectively. Also a pressure induced polymorphic phase transition is seen using DFT functionals and the same was not observed with DFT-D2 method. The equation of state and compressibility of both polymorphic phases were investigated. Electronic structure and optical properties were calculated using full potential linearized augmented plane wave method within the Tran-Blaha modified Becke-Johnson potential. The calculated electronic structure shows that $\alpha$, $\beta$ phases are indirect band gap insulators with a band gap values of 3.51 and 4.43 eV, respectively. The nature of chemical bonding is analyzed through the charge density plots and partial density of states. Optical anisotropy, electric-dipole transitions and photo sensitivity to light of the polymorphs are analyzed from the calculated optical spectra. Overall, the present study provides an early indication to experimentalists to avoid the formation of unstable $\beta$-form of AgCNO.       
\end{abstract}

\maketitle
\section {Introduction}
Inorganic fulminates are primary explosives and they find applications as initiators for secondary explosives. They are iso electronic with the corresponding azides, cyanates and cyanamides \cite{iqbal1}. Silver fulminate (AgCNO) belongs to the class of inorganic fulminates, which is an effective detonator and also found to be as effective as lead azide (Pb(N$_3$)$_2$) in its pure form (Ex. 60 mg of AgCNO is sufficient to get maximum output from the Hexahydro 1,3,5-trinitro-1,3,5-triazine (RDX)) \cite{collins}. It is about fifteen times as efficient as the mercury fulminate (Hg(CNO)$_2$) for exploding 2,4,6-tri-nitro-phenyl-methyl-nitroamine and 30\% more efficient for exploding tri-nitro-xylene \cite{davis}. It is known to be a sensitive explosive for a long time with good detonating properties \cite{davis,collins}, the average detonation velocity of AgCNO sample of thickness $\sim$0.5 mm is in the order of 1700 ms$^{-1}$, when ignited by a hot wire\cite{bowden}. AgCNO detonators has been used in the Italian Navy \cite{urbanski}. It finds some applications (when small quantity is used) in pyrotechnics, fire works, and toy pistols \cite{springer}. The decomposition mechanism of AgCNO is as follows: 2AgCNO $\rightarrow$ 2Ag + 2CO + N$_2$; and the production of two gases CO and N$_2$ give distinct noise when it is used in Christmas crackers\cite{jackson}. Moreover, the initiating propeties of various fulminates were examined by Martin and wohler \cite{martin1,martin2,martin3}, and they found that the Ag, Cd and Cu fulminates have stronger initiating power than Hg(CNO)$_2$. However, Hg(CNO)$_2$ stands first which is now being widely used as standard primary explosive, but it was largely replaced by Pb(N$_3$)$_2$ \cite{miles} due to its poor stability and toxicity. The Pb(N$_3$)$_2$ was found to have better thermal stability and insensitive compared to the fulminates. Measured figure of insensitiveness (F of I) to impact for Hg(CNO)$_2$ (10) $\textless$ AgCNO (22) $\textless$ Pb(N$_3$)$_2$ (30) and temperature of ingnition (T of I) for AgCNO (170$^o$C) $\textless$ Hg(CNO)$_2$ (210$^o$C) $\textless$ Pb(N$_3$)$_2$ ($\textgreater$ 300$^o$C) [Ref \onlinecite{collins} and references there in]. However, it has been reported that AgCNO crystals stored under water about 40 years ago exhibit similar properties as like fresh crystals with an exception that the white crystals transform to mouse-gray colour due to long exposure to light and these crystals possess non hygroscopic nature \cite{peter1,collins}.           

\par AgCNO was first prepared by Brugnatelli \cite{brugnatelli} (1798) and later recognised by Gay-Lussac \cite{gay} (1824). In 1959, Singh\cite{singh1} reported the crystal structure of AgCNO to be body centred orthorhombic structure having space group $Ibam$ with lattice constants a = 3.88 $\AA$, b = 11.20 $\AA$ and c = 6.04 $\AA$, and Z = 4. The accuarte crystal symmetry and formation of two ploymorphic phases of AgCNO was first pointed out by Pandey \cite{pandey} and Britton $\emph{et al}$ \cite{britton1} (1965). Further, the redetermined crystal structures of the two polymorphic phases are reported as orthorhombic (\emph{Cmcm}) phase by Barrick $\emph{et al}$ \cite{barrick} (1979), with unit-cell parameters a = 3.88 $\AA$, b = 10.752 $\AA$ and c = 5.804 $\AA$, and Z = 4, and trigonal (\emph{R$\bar{3}$}) phase by Britton \cite{britton2} (1991), a = 9.087$\AA$, $\alpha$ = 115.73$^{\circ}$ and Z = 6. The differential thermal analysis studies \cite{boddington} indicate no evidence for temperature induced phase changes between 20$^o$C and 145$^o$C. To the best of our knowledge, there are no studies to explore the relative stability under pressure and temperature, electronic structure, and chemical bonding of the two polymorphic phases in the literature from both theoretical and experimental point of view for the energetic solid AgCNO. Therefore, we systematically investigated the above mentioned properties for AgCNO using first principles calculations based on Density Functional Theory (DFT). The rest of the article is organized as follows, in section II, we briefly describe the methodology of our calculations. In section III, the polymorphic phase stability, electronic struture, chemical bonding, and optical properties of AgCNO are discussed. Finally, in section IV, we summarize the results, which concludes our paper.

\section{Methodology of calculations}
The polymorphic phases of AgCNO crystals were studied using periodic DFT calculations performed with two different codes such as Plane Wave Self Consistent Field (PWSCF) \cite{Giannozzi} and Cambridge series of total energy package (CASTEP) \cite{payne}, which are based on planewave pseudopotential (PW-PP) approach. We have used Vanderbilt type ultrasoft pseudo potentials\cite{vander} to treat ion-electron interactions, while electron-electron interactions are treated with the generalized gradient approximation (GGA) by Perdew-Burke-Ernzerhof (PBE)\cite{burke} and Perdew
and Wang (PW91)\cite{wang}. The Broyden-Fletcher-Goldfarb-Shanno (BFGS) minimization scheme \cite{almlof} was used in geometry optimization. The plane wave cutoff energy was set to 60 Ry and a k-point grid of 7$\times$7$\times$4 and 5$\times$5$\times$5 for \emph{Cmcm} and \emph{R$\bar{3}$} phases, respectively using Monkhorst-Pack grid scheme \cite{monkhorst}. The self-consistent energy convergence accuracy was set to 1.0$\times$10$^{-6}$ eV/atom and the force to be 1.0$\times$10$^{-4}$ eV/A. 
\par The standard LDA/GGA functionals are unabale to describe the non-covalent interactions in molecular crystals. On the other hand, semiemipirical dispersion correction methods such as the Ortmann, Bechstedt, and Schmidt (DFT-OBS)\cite{Ortmann} correction to PW91, as well as the Tkatchenko and Scheffler (DFT-TS)\cite{Tkatchenko} and Grimme (DFT-D2)\cite{Grimme} corrections to PBE were employed to handle vdW interactions. These methods are found to improve the results to a greater extent. Here the total energy after inclusion of vdW correction is given by
\begin{equation}
E_{DFT+D} = E_{DFT} + E_{disp}
\end{equation}
Where E$_{DFT}$ is the self-consistent Kohn-Sham energy, E$_{disp}$ is empirical dispersion correction given by 
\begin{equation}
E_{disp} = -S_6\sum\limits_{i \textless j}\frac{C_{ij}}{R_{ij}^6}f_{damp}(R_{ij})
\end{equation}
Where S$_6$ is global scaling factor that only depends on the density functional used. C$_{ij}$ denotes the dispersion coefficient for the pair of i$^{th}$ and j$^{th}$ atoms that depends only on the chemical species, and R$_{ij}$ is an interatomic distance. f$_{damp}$ = $\frac{1}{1+e^{-d(R_{ij}/R_0-1)}}$ is a damping function which is necessary to avoid divergence for small values of R$_{ij}$ and R$_0$ is the sum of atomic vdW radii. The DFT-D2 method has been applied to several molecular systems such as secondary explosives\cite{Zheng, Sorescu,kondaiah1, kondaiah2} and inorganic azides\cite{ramesh1, ramesh2,ramesh3}, and these results demostrated that dispersion corrections are important in determining the ground state properties. Hence, we attempted to study the phase stability of AgCNO using the DFT-D2 method and the results are discussed in the following section. 

The thermodynamic stabilities were investigated by comparing the free energy of the polymorphs. In order to predict the most stable form of AgCNO, we have calculated the Gibbs free energy of both polymorphic phases. The Gibbs free energy for any material at a given temperature is as follows.
\begin{equation}
G(P, T) = F(V, T) + PV = F_{vib} + F_{perfect} + PV
\end{equation}
Where F, P, T, and V are the Helmoltz free energy, pressure, absolute temperature and volume, respectively. F(V,T) is summation of vibrational free energy and perfect lattice energy \emph{i.e} F(V, T) = F$_{vib}$ + F$_{perfect}$. The vibrational free energy F$_{vib}$ is calculated within the harmonic approximation and is given by
\begin{equation}
 F_{vib} = K_BT\sum\limits_{i}\Big{(}\frac{h\omega_i}{2K_BT} + ln(1-e^{\frac{-h\omega_i}{K_BT}})\Big{)}
\end{equation}
Where $\omega_i$ is the phonon frequency, $h$ is the Planck constant and $K_B$ is the Boltzmann constant. Where F$_{perfect}$ = E$_0$ + E$_{el}$ - TS$_{el}$, E$_0$ and E$_{el}$- TS$_{el}$ are the contributions from the lattice and electronic excitations, respectively. If the electronic excitations and pressure changes are neglected then the Gibbs free energy can be calculated by G =  E$_{0}$ + F$_{vib}$.  

\section{Results and discussion}
\subsection{Crystal structure and phase stability of polymorphs}
Extreme sensitivity, poor stability and high cost of Silver as a raw material prohibited AgCNO in commercial or military priming and detonating devices \cite{taylor1,taylor2} but still it finds applications in pyrotechincs, fire works, toy pistols and in the navy. However, the relative stability and  necessary conditions for the formation of two polymorphic phases are still unknown for this material. As discussed in section I, AgCNO exists in two polymorphic phases at ambient conditions, and it is noticed that except the crystal structures \cite{britton1,barrick,britton2} of both phases most of the physical properties are not well understood from experimental and theoretical prospective. Hence we take an opportunity to investigate the relative thermodynamic phase stability of the polymorphs. 

\par
\begin{figure*}
\centering
\includegraphics[height = 3.5in, width=7.0in]{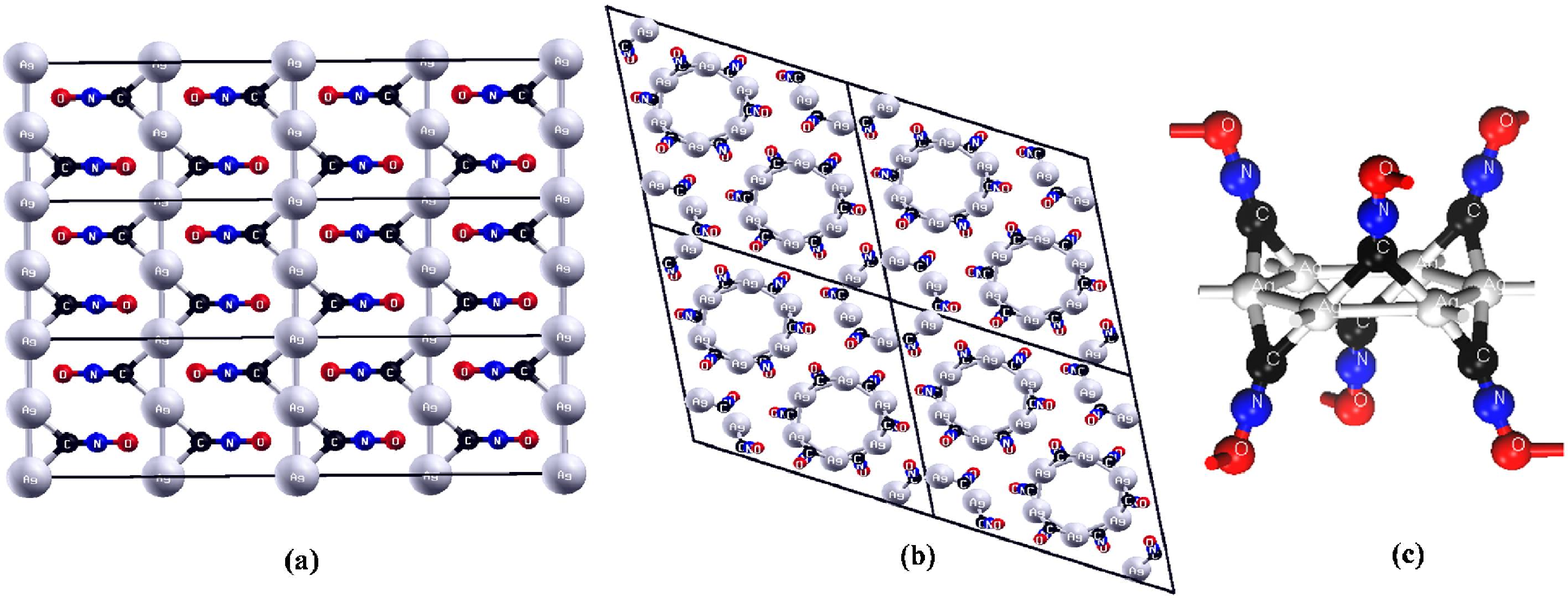}
\caption{(Color online) Crystal structures of a) $\alpha$-phase (2x3 super cell) along yz-plane, b) $\beta$-phase (2x2 super cell) along xy-plane and c) single molecule of $\beta$-phase.}
\label{cmcm}
\end{figure*}

 As a first step, we have performed full structural optimization by taking experimental data as a input.\cite{barrick,britton2} The lattice parameters and fractional co-ordinates of both the phases are allowed to relax to get the most stable configurations at ambient pressure. The optimized crystal structures of orthorhombic (\emph{Cmcm}) and rhombohedral (\emph{R$\bar{3}$}) phases are shown in Fig. \ref{cmcm}. The crystal structures of two polymorphic phases mainly differ by planar zigzag chains (orthorhombic phase in yz plane) and cyclic hexamers (rhombohedral phase in xy plane)\cite{britton1} (see Fig. \ref{cmcm}) and these AgCNO molecules are bind through weak vdW forces within the unitcell. We have calculated the structural properties and their pressure dependence within standard DFT functionals using PWSCF and CASTEP codes. The obtained volumes for \emph{Cmcm} and \emph{R$\bar{3}$} phases are overestimated by 22.5$\%$ and 13.5$\%$ with PBE-GGA, respectively. This clearly represents that PBE-GGA functional is inadequate to predict the ground state properties of this energetic molecular solid. Therefore, we have used various dispersion corrected (DFT-D) methods to capture vdW interactions to reproduce the ground state properties comparable with the experiments.\cite{britton1,barrick,britton2} The obtained volumes are overestimated by about 12.6$\%$, 7.8$\%$ using DFT-OBS; 3.7$\%$, 4.2$\%$ using DFT-TS; and 2.2$\%,$ 4.5$\%$ using DFT-D2 methods for \emph{Cmcm}, \emph{R$\bar{3}$} phases, respectively. Among the three DFT-D methods, DFT-TS and DFT-D2 methods work well for the molecular solid AgCNO, still there are dicrepancies between the theoretical values at 0 K and experimental data at 298 K\cite{britton1,barrick,britton2}. The order of discrepancies about $\sim$3-4$\%$ are previously reported for seconadry explosive molecular crystals with DFT-D methods at 0 K\cite{landerville,vashishta1,Sorescu} and these errors were further reduced by inclusion of thermal effects in the calculations\cite{landerville,vashishta2} and the obtained results are in excellent agreement with the experimental data. In the present study, the equilibrium volumes are overestimated within static DFT-D2, if we include the temperature effects there will be an increment in the volumes due to thermal expansion which leads to large dicrepancies over static DFT-D2 results unless the polymorphs have negative thermal expansion. The calculated ground state lattice parameters, unitcell volume, and density of both polymorphic phases using various dispersion corrected DFT functionals are compared with experimental data and are presented in Table I. The obatined ground state properties with and without dispersion correction methods using PWSCF and CASTEP are in good accord with each other (see Table I). In general, the energy difference between different polymorphic forms are mostly in the order of 0-10 KJ/mole.\cite{Baraga} The DFT-D2 method is successful in predicting the polymorphs of Benzamide (P1 and P3) with an energy difference by 1.9 KJ/mole.\cite{Ectors} Hence, we have used DFT-D2 method for further calculations. The obtained total energy difference for \emph{Cmcm} phase lowered by $\sim$6 KJ/mole per molecule than \emph{R$\bar{3}$} phase which implies that \emph{Cmcm} phase found to be the thermodynamic ground state at ambient pressure. Britton et al \cite{britton1} proposed that \emph{Cmcm} is stable than \emph{R$\bar{3}$} phase due to the presence of linear form of CNO anions in their crystal structures.\cite{iqbal2} The predicted behavior with DFT-D2 method are in good agreement with the experimental observations\cite{britton1,barrick,britton2,iqbal2} and this clearly shows the success of DFT-D2 method in predicting the relative phase stability of the ploymorphic phases of AgCNO.   

\subsection{Pressure and temperature effects on the polymorphs}
\begin{figure*}
\centering
\includegraphics[height = 2.5in, width=6.0in]{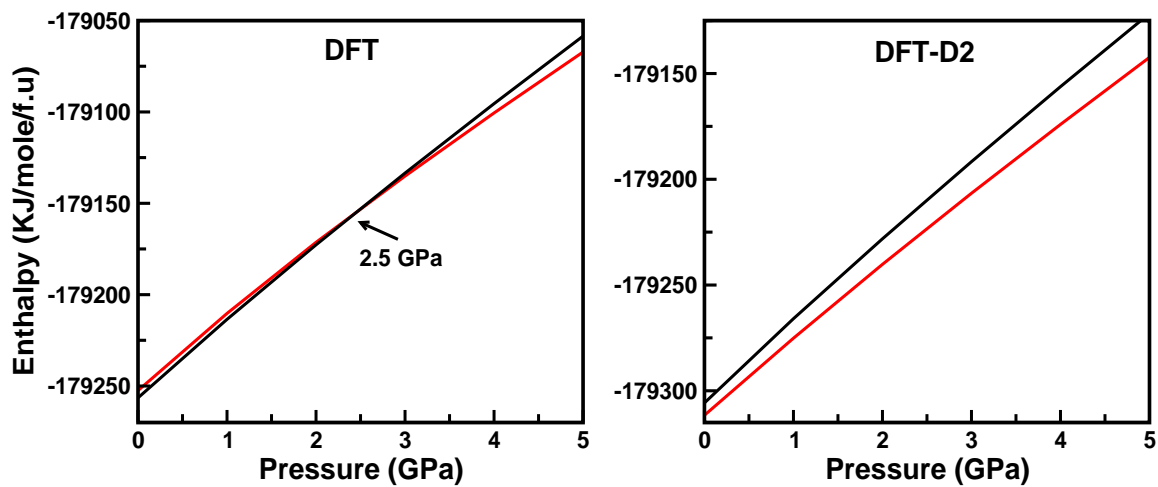}
\caption{(Color online) Calculated enthalpy as a function of pressure for $\alpha$- and $\beta$- polymorphic phases of AgCNO with (DFT-D2) and without (PBE-GGA) dispersion correction method} 
\label{PT}
\end{figure*}

The investigation of energetic materials at extreme conditions is a challenging task because of their sensitivity, complex chemical behavior and risk of decomposition. Several accidents occured during the preparation of AgCNO\cite{collins} and hence it is very difficult to perform experiments without any prior knowledge about this kind of energetic materials. Therefore, theoretical modeling and simulations are efficient tools to predict the physical and chemical properties of complex energetic solids at extreme conditions and also very useful in predicting results a priori to the experimentalists. DFT is a powerful tool in predicting the behavior of complex solid state systems at extreme conditions. Hence, we attempted to study the relative phase stability and possible polymorphic structural transition in AgCNO under hydrostatic pressure upto 5 GPa with a step size of 0.5 GPa. As illustrated in Fig. \ref{PT}, The calculated enthalpy curves as a function of pressure using standard DFT functionals show that \emph{R$\bar{3}$} phase undergoes a structural ploymorphic phase transition to \emph{Cmcm} phase at about 2.5 GPa within GGA using PWSCF and the corresponding transition pressures obtained from CASTEP code is 2.7 GPa (see Fig. 1 of supplementary material\cite{support}). Overall, we observed a similar trends using both of the approaches with small deviations in the calculated transition pressures. Above the transition pressures the enthalpy of the \emph{Cmcm} phase is lower than that of the \emph{R$\bar{3}$} phase, indicating that \emph{Cmcm} structure becomes stable. The \emph{Cmcm} phase is stable in the pressure range from transition pressure to 5 GPa, which is the highest pressure we accomplished in the present work. The enthalpy (total energy) difference at 0 GPa is $\sim$ 6 KJ/mole which increases with pressure and it is found to be $\sim$ 21 KJ/mole at 5 GPa within DFT-D2 method. In contrast to LDA/GGA functionals, the DFT-D2 method using both of the approaches (PWSCF and CASTEP) reveals that there is no pressure induced ploymorphic phase transition seen in AgCNO and the \emph{Cmcm} phase is found to be the most stable form of AgCNO over studied pressure range using both PW-PP approaches (see Fig. \ref{PT}). Hitherto, we represent \emph{Cmcm} and \emph{R$\bar{3}$} phases as $\alpha$ and $\beta$-AgCNO, respectively. Our recent high pressure study on KClO$_3$\cite{kondal} also revealed that DFT-D2 method is good enough for calculating the transition pressures for molecular solids in opposition to standard LDA/GGA functionals. As shown in Fig. \ref{PTE} (see Fig. 2 of supplementary material\cite{support}), the calculated total energies ploted as a function of pressure demonstrates that the DFT-D2 functional predict the $\alpha$-phase to be the most stable form of AgCNO over studied pressure range in contrast to usual LDA/GGA functionals. 

In addition, the contribution of lattice vibration needs to be considered at elevated temperatures. We have used norm-conserving pseudo potentials\cite{Troullier} for calculating the Gibbs free energy as they are well suited for phonon calculations as implemented in CASTEP code. Fig. \ref{GT} shows the Gibbs free energy as a function of temperature in the range of 0-450 K within DFT-D2 method. The difference in Gibbs free energy ($\Delta$G) is $\sim$ 7 KJ/mole at low temperature region and it decreases monotonically with increase in temperature between the two $\alpha$- and $\beta$-polymorphs of AgCNO. However, $\alpha$-AgCNO remains with the lowest free energy throughout the investigated temperature range, despite the contribution of lattice vibration tending to decrease the $\Delta$G gap at elevated temperatures. We have not observed any temperature driven phase transition between two polymorphs over studied temperature range and this is consistent with results of differential thermal analysis\cite{boddington}. Gibbs free energies at temperatures beyond 450 K are not plotted because $\alpha$-AgCNO phase typically decompose at higher temperatures between 393-450 K\cite{boddington}. From our present study, we confirmed that $\alpha$-phase is the most stable polymorph of AgCNO under studied pressure and temperature range.
                
\subsection{Equation of state and compressibilities of the polymorphs}
To study the effect of hydrostatic pressure on crystal structures of both the polymorphs, we have used variable cell optimization technique. The calculated volumes decrease monotonically as a function of pressure using standard LDA/GGA functionals as well as with DFT-D2 method are shown in Fig. \ref{PV} (see Fig. 3 of supplementary material\cite{support}) and the obtained equilibrium bulk moduli for $\alpha$ and $\beta$-AgCNO are found to be 20.0 (19.4) GPa and 17.8 (19.1) GPa within DFT-D2 method using PWSCF (CASTEP) by fitting pressure-volume data to second order Birch-Murnaghan equation of state.\cite{murnaghan} In order to understand the behavior of unit-cell parameters and their relative compressibilities under compression, we have presented the lattice constants as a function of pressure in Fig. \ref{abc}. The calculated lattice constant 'a' decreases monotonically with pressure whereas the b and c lattice constants are decreasing non-monotonically in the pressure range of 2.5 to 5 GPa. The computed lattice constants are very well reproduced with fifth-degree polynamials and the fitting pressure coefficients are given in Table II as previosly reported for explosive nitrate ester 1 (NEST-1).\cite{white} As illustrated in Fig. \ref{abc} and from the calculated first order pressure coefficients, the c-axis and b-axis are the most and least compressible, respectively for $\alpha$-AgCNO while rhombohedral lattice constant (a) and angle ($\alpha$) are decreasing with distinct pressure coefficients under pressure for $\beta$-AgCNO. This clearly indicates that the polymorphs behave anisotropically under the hydrostatic pressure.  Further, to understand the relative compressibilities of both the polymorphic phases bondlengths and bond angles as a function of hydrostatic pressure are plotted within DFT-D2 method. As shown in Fig. \ref{BL}, Ag-O bond is more compressible in both the phases along a-crystallographic direction whereas Ag-Ag is the most compressible in $\beta$-AgCNO. The bonds such as Ag-C, C-N, and N-O show similar and less compressible nature over studied pressure range due to strong covalent bonding between Ag and C as well as within the CNO anion molecule and this can be clearly understood from their electronic structure. Similarly the corresponding bond angles Ag-C-Ag decrease because of the orientation of Ag-C bonds due to high compressibility nature of Ag-Ag bonds (see Fig. \ref{BL}) whereas Ag-C-N bond angle increases under compression (see Fig. \ref{BA}). Overall, we observe that Ag-C, C-N, and N-O bonds are stiffer, while Ag-Ag and Ag-O are more compressible under the application of hydrostatic pressure. Further, this compressibility behavior can be clearly understand by analyzing the nature of chemical bonding in AgCNO and hence we have investigated the electronic struture and chemical bonding of both the polymorphic phases in the following section.    

\subsection{Electronic structure and chemical bonding}
Inorganic fulminates are isoelectronic with azides, cynates, and cynamides. Iqbal and his co-workers \cite{iqbal1} explained that these fulminates are found to be the most sensitive explosives to shock and heat among the energetic materials such as azides, cynates, thiocynates, and cynamides due to asymmetric charge distribution in their structures. Also, the heavy metal complex salts are unstable than alkali metal salts because of the asymmetric inter ionic distances \cite{iqbal2}. In order to understand the macroscopic energetic behavior of these sensitive explosive materials, it is necessary to understand the electronic structure, and chemical bonding of these materials at the microscopic level. Iqbal et al \cite{iqbal1} carried out a detailed experimental study on the electronic structure and stability of the inorganic fulminates, which reveals that sodium, potasium, and thallous fulminates to be ionic salts whereas silver and mercury salts are covalent in nature and this will be reflected in the order of stability of the fulminate salts. However, the electronic structure and bonding properties of the polymorphs of silver fulminate salts are not reported in the literature. Therefore in the present study we made a detailed analysis of electronic structure and chemical bonding of both the polymorphic phases at ambient pressure. 

\par It is well known fact that the standard LDA/GGA functionals severely underestimate the band gaps for semiconductors and insulators. In order to get relaiable energy band gaps several methods have been proposed such as LDA+U, LDA+DMFT, hybrid fuctionals, and GW approximation. Unfortunately these methods are computationally very expensive.  However, Tran and Blaha modified Becke Johnson (TB-mBJ) potential \cite{peter2} is computatinally less expensive method, which provides accurate energy band gaps as comparable with more sophisticated methods within the Kohn-Sham frame work. Recently, Koller \emph{et al}\cite{koller1,koller2} reported the merits and limitations of the TB-mBJ functional. So far, several groups used the TB-mBJ potential for the calculation of electronic structure and optical properties of different kinds of materials \cite{singh2,camargo,jiang} and confirmed that bandgaps obtained using TB-mBJ potential are improved when compared to LDA/GGA functional for a wide range of materials (see Fig. 1 from Ref. \onlinecite{dixit}). The accurate prediction of band gaps using TB-mBJ functional for diverse materials motivated us to use this functional for high energetic materials. This semi local functional is implemented through WIEN2k package \cite{blaha}. Since all these energetic materials are semiconductors or insulators, we have used the TB-mBJ functional to calculate energy band gap values for the two ploymorphs. We have used experimental crystal structures\cite{barrick,britton2} to calculate the electronic structure and optical properties of the both polymorphs at ambient pressure. The calculated band gaps are found to be 3.51 and 4.43 eV for $\alpha$ and $\beta$-phases of AgCNO, respectively and the respective band gaps using LDA functional are 2.00, 2.85 eV. The obtained TB-mBJ band gap of 3.51 eV for $\alpha$-AgCNO is closely comparable with the measured optical energy gap of 4.0 eV\cite{iqbal1} over LDA functional. Also, the calculated TB-mBJ band structures show that $\alpha$ and $\beta$-AgCNO phases are indirect band gap insulators along S-($\Gamma$-Z) and T-(F-$\Gamma$) high symmetry directions of the Brillouin zone as shown in Fig. \ref{BS}. 

\par Further, the nature of chemical bonding in the two polymorphic phases was investigated by examining the total and partial density of states (DOS) as illustrated in Fig. \ref{DOS}. The relative phase stability of the $\alpha$ and $\beta$ polymorphs of AgCNO can also be explained on the basis of the DOS. As shown in Fig. \ref{DOS}, when compared to $\beta$-AgCNO the peaks of the DOS in the valence bands of $\alpha$-AgCNO has a tendency to shift towards lower energy, indicating better stability of the $\alpha$ polymorphic phase. Also, it can be seen that total DOS of both phases exhibit some similar features, hence we have analyzed the bonding based on PDOS of $\alpha$-AgCNO. It is clearly visualized from the DOS that AgCNO polymorphs have molecular character, which arises from a strong overlap contributions of partial DOS of Silver and Fulminate anions in the valence and conduction bands  and this can be commonly seen in molecular crystalline solids. The conduction band is mainly due to $p$-states of C, N, O and $s,p,d$-states of metal (Ag) atoms. The lowest lying states about -9 eV are due to hybridized C-$s$ and $s$, $p$ states of N and O atoms. The states from -5 to -7 eV are due to hybridization of Ag-$d$ and $p$-sates of C, N and O atoms. Also, the top of the valence band is mainly dominated by O-$p$ and Ag-$d$ states and there is a charge sharing between C-$p$ and Ag-$d$ states representing the covalent nature of Ag-C bond. The nature of chemical bond between two atoms can be predicted from their electronegativity difference.  When the differences are greater or equal to 1.7 (ionic) and less than 1.7 (covalent), if the difference is greater than 0.5 the covalent bond has some degree of polarity. According to Pauling scale the electronegetivity values Ag (1.9), C (2.5), N (3.0), and O (3.4) shows that there exists a covalent bond between N-O, C-N and Ag-C bonds. The electronegetivity difference in N-O (0.4), C-N (0.5) and Ag-C (0.6) exhibits strong and polar covalent nature of N-O, C-N and Ag-C bonds, respectively. The N-O, C-N and Ag-C bonds show less compressibility behavior with increasing pressure (see Fig. \ref{BL}), this is due to strong hybridization between Ag-$d$ and $s$, $p$-states of C, N, and O atoms leads to strong covalent character. Further, this can be clearly understood from electronic charge density plots which are used for accurate description of chemical bonds.\cite{Hoffman} The calculated valence charge density using TB-mBJ functional for both the polymorphs are as shown in Fig. \ref{charge}. It shows anisotropic bonding interactions and the charge cloud is distributed within the CNO molecule indicating covalent character. Overall, the C, N, and O atoms are covalenlty bonded within CNO anion and the metal atom is also covalently bonded with CNO group through C atom. X-ray electron spectroscopy\cite{Colton} study on inorganic azides reveals that Heavy Metal azides (HMAs) are more covalent than Alkali metal azides (AMAs), implying that HMAs are more sensitive than AMAs and find applications as intiators for secondary explosives. Therefore, the presence of covalent bonding in AgCNO makes it more sensitive than the above mentioned ionic inorganic fulminates and it can be used as an initiator as HMAs.

\subsection{Optical properties}
Energetic materials become unstable and they undergo photochemical decomposition by the action of light. Hence it is interesting to study the optical properties of these materials to understand the decomposition mechanisms. Electronic structure calculations could provide an information about the nature and location of interband transitions in crystals. The complex dielectric function $\epsilon(\omega)$ = $\epsilon_1(\omega)$ + $i\epsilon_2(\omega)$ can be used to describe the linear response of the system to electromagnetic radiation which is related to the interaction of photons with electrons. The imaginary part of dielectric function $\epsilon_2(\omega)$ is obtained from the momentum matrix elements between the occupied and unoccupied wavefunctions within selection rules. The optical properties of AgCNO polymorphs were calculated using TB-mBJ functional at a denser k-mesh of 20 $\times$ 20 $\times$ 12 and 17 $\times$ 17 $\times$ 17 for $\alpha$- and $\beta$- phases, respectively. The AgCNO polymorphs crystallize in the orthorhombic and hexgonal symmetry, and this allow non zero components of the dielectric tensors three for $\alpha$ and two for $\beta$ phase along [100], [010] and [001] directions. Fig. \ref{epsilon} shows the real $\epsilon_1(\omega)$ and imaginary $\epsilon_2(\omega)$ part of the dielectric function as a function of photon energy for both the polymorphs. The peaks in $\epsilon_2(\omega)$ mainly arises due to electric-dipole transitions between valence and conduction bands. The major peaks in $\epsilon_2(\omega)$ at 4.57 eV (along 100), 5.94 eV (along 010), 5.06 eV (along 001) in $\alpha$-phase and 5.45 eV (along 100), 6.13 eV (along 001) in $\beta$-phase arise due to interband transitions from Ag($4d$) $\rightarrow$ N($2p$)-states. It can be clearly seen from DOS, the top of valnce band is mainly dominated by $4d$-states of Silver atom (see Fig. \ref{DOS}) in both the polymorphs, which is similar to the case of Silver azide (AgN$_3$)\cite{pradeep}. As discussed in the section I, AgCNO is iso-electronic and it consists of same chemical species as AgN$_3$ in addition to C and O atoms, hence one can expect that major optical transitions arise from Ag($4d$) $\rightarrow$ N($2p$)-states in both energetic materials AgCNO and AgN$_3$\cite{pradeep}. 

\par The real part $\epsilon_1(\omega)$ of can be derived from $\epsilon_2(\omega)$ using the Kramer-Kronig relations. The calculated real static dielectric constant along all the crystallographic directions are 2.24, 5.01, 2.78 for $\alpha$-phase and 2.59, 4.38 for $\beta$-phase. The static refractive indices calculated using the dielectric function n = $\sqrt{\epsilon(0)}$ are given by n$_{100}$ = 1.50, n$_{010}$ = 2.24, n$_{001}$ = 1.67 and n$_{100}$ = 1.61, n$_{001}$ = 2.09 for $\alpha$ and $\beta$-phases of AgCNO, respectively. The obtained refractive indices are distinct in all the crystallographic directions, which indicates the anisotropy of the $\alpha$ and $\beta$ polymorphs. Using $\epsilon_1$, $\epsilon_2$ one can dervive the optical constants such as absorption and photoconductivity of the materials. The calculated absorption spectra is shown in Fig. \ref{absorp} and absorption starts after the energy 3.51 eV in $\alpha$-phase, 4.43 eV in $\beta$-phase, which is energy band gap between the valence band maximum and conduction band minimum. The first absorption peaks along three (100, 010 and 001) directions are at 4.5 eV, 7.7 eV, 7.9 eV and the corresponding absorption coefficients are found to be 1.1 $\times$ 10$^7$m$^{-1}$, 1.4 $\times$ 10$^8$m$^{-1}$, 3.9 $\times$ 10$^8$m$^{-1}$ for $\alpha$-phase, while the same are 1.3 $\times$ 10$^8$m$^{-1}$ at 6.8 eV and 4.0 $\times$ 10$^7$m$^{-1}$ at 7.2 eV along two (100 and 001) directions for $\beta$-phase. Photoconductivity is due to the increase in the number of free carriers when photons are absorbed. The calculated photoconductivity shows (see Fig. \ref{absorp}) a wide photocurrent response in the absorption region of 3.5-35 eV and 4.4-35 eV in  $\alpha$ and $\beta$ phases of AgCNO, respectively. overall, the energetic polymorphs show a strong anisotropic and wide range of absorption (absorption coefficients $\sim$ 10$^7$m$^{-1}$). This results indicate the possible decomposition of AgCNO into Ag, CO and N$_2$ under the action of ultraviolet (UV) light. Therefore AgCNO polymorphs decompose under the action of UV light and they may explode due to photochemical decomposition.

\section{CONCLUSIONS}
In summary, \emph{ab-initio} calculations have been performed to investigate the relative phase stability, structural transition and compressbilities of both polymorphic orthorhombic (\emph{Cmcm}) and trigonal (\emph{R$\bar{3}$}) phases of AgCNO using PW-PP approaches. The standard LDA/GGA functionals are inadequate to predict the relative phase stability and ground state properties of these polymorpic phases. From our total energy calculations using DFT-D2 method, \emph{Cmcm} phase was found to be preferred thermodynamic equilibrium phase under pressure and temperature range and hence we confirmed \emph{Cmcm} and \emph{R$\bar{3}$} phases as $\alpha$ and $\beta$-AgCNO, respectively. The present study also reveals that there is no structural phase transition observed using DFT-D2 method whereas we observe a pressure induced polymorphic phase transition from $\beta$ $\rightarrow$ $\alpha$ at about 2.5 (2.7) GPa within GGA using PWSCF (CASTEP) package. Overall, the phase stability and ground state properties obtained from both the PW-PP approaches are in good accord with each other for both of the polymorphs. We have also calculated the electronic structure and density of states of both polymorphic phases using TB-mBJ functional at ambient pressure. The calculated electronic band structure show that $\alpha$, $\beta$ phases are indirect band gap insulators with a band gap values of 3.51 and 4.43 eV, respectively. The detailed analysis of electronic charge density distribution and partial density of states reveal covalent bonding is predominant in the energetic AgCNO polymorphs. The major peaks in $\epsilon_2(\omega)$ are due to interband transitions between Ag($4d$) $\rightarrow$ N($2p$)-states. The calculated absorption spectra reveal that the polymorphs show considerable anisotropy and decompose under the action of UV light. Finally, the present study suggests that the $\alpha$-form is stable and it can be used in different applications and $\beta$-form can be avoided by careful control of the pressure or temperature in the manufacturing process.  

\section{Acknowledgments}
Authors would like to thank Defence Research and Development Organisation (DRDO) through ACRHEM for the financial support under grant No. DRDO/02/0201/2011/00060:ACREHM-PHASE-II, and the CMSD, University of Hyderabad, for providing computational facilities. NYK acknowledges Dr. V. Kanchana, Department of Physics, Indian Institute of Technology Hyderabad for critical reading of the manuscript.
 \\
$^*$\emph{Author for Correspondence, E-mail: gvsp@uohyd.ernet.in}

{\pagestyle{empty}

\begin{table*}
\caption{Calculated (PWSCF and CASTEP) ground state lattice parameters (a, b, c, in $\AA$), crystallographic angle ($\alpha$, in $^{\circ}$), volume (V, in $\AA^3$), and density ($\rho$, in gr/cc) of $\alpha$- and $\beta$- polymorphic phases of AgCNO using different dispersion corrected DFT functionals along with experimental data.}
\begin{ruledtabular}
\begin{tabular}{ccccccccccc}
        &           &          &             &   $\alpha$-AgCNO &  &           &        & $\beta$-AgCNO   &    &         \\ \hline
Package &  Method   &    a     &    b        &     c     &   V     &  $\rho$   &   a    &  $\alpha$  &   V  &  $\rho$  \\ \hline
PWSCF   &   PBE     &  4.701   &   10.765    &  5.869  &  297.00   &  3.35    & 9.571  &   115.99    &   442.94   &  3.37    \\ 
        &  DFT-D2   &  3.411   &   11.467    &  6.327  &  247.50   &  4.02    & 9.264  &   115.84    &   409.03   &  3.65    \\
CASTEP  &  PBE      &  4.773   &   10.743    &  5.847   & 299.80   &  3.32    & 9.550  &   116.06    &   436.64   &  3.42   \\
        &  DFT-D2   &  3.416   &   11.436    &  6.340   & 247.69   &  4.02    & 9.264  &   115.86    &   407.76   &  3.66   \\
        &  DFT-TS   &  3.883   &   11.009    &  5.872   & 251.05   &  3.97    & 9.223  &   115.76    &   407.04   &  3.67   \\
        &  DFT-OBS  &  4.352   &   10.749    &  5.826   & 272.54   &  3.65    & 9.379  &   115.90    &   421.24   &  3.54   \\ 
  &  Expt.$^{a,b}$  &  3.880   &   10.752    &  5.804   & 242.13   &  4.11    & 9.087  &   115.73    &   390.58   &  3.82  \\                       
\end{tabular}
\end{ruledtabular}
$^a$ Ref.\cite{barrick}, $^b$ Ref.\cite{britton2}
\end{table*}

\begin{table*}
\caption{Fifth degree polynomial fits for pressure dependence of lattice parameters for $\alpha$- and $\beta$- polymorphic phases of AgCNO using PWSCF within DFT-D2 method. Each constant C$_n$(n = 0-5) has units of  $\AA$/(GPa)$^n$.}
\begin{ruledtabular}
\begin{tabular}{cccccc}
                             &           &   $\alpha$-AgCNO   &          & \hspace{0.2in} $\beta$-AgCNO  \\ \hline
   Parameter                 &    a      &    b     &   c     &    a     &  $\alpha$       \\ \hline
    C$_0$                    &  3.4114   &  11.467  & 6.3271  &   9.2677 &   115.84         \\
C$_1$ ($\times$ 10$^{-1}$)   & -0.693    & -0.446   & -1.161  &  -2.686  &  -2.263          \\
C$_2$ ($\times$ 10$^{-2}$)   &  2.085    & -6.131   & -3.011  &  -3.928  &  -9.700          \\
C$_3$ ($\times$ 10$^{-2}$)   & -0.697    &  3.032   &  1.509  &   4.810  &   7.062          \\         
C$_4$ ($\times$ 10$^{-2}$)   &  0.109    & -0.553   & -0.242  &  -1.201  &  -1.645          \\
C$_5$ ($\times$ 10$^{-3}$)   & -0.065    &  0.324   &  0.141  &   0.970  &   1.294          \\                    
\end{tabular}
\end{ruledtabular}
\end{table*}

}

\begin{figure*}
\centering
\includegraphics[height = 3.0in, width=4.0in]{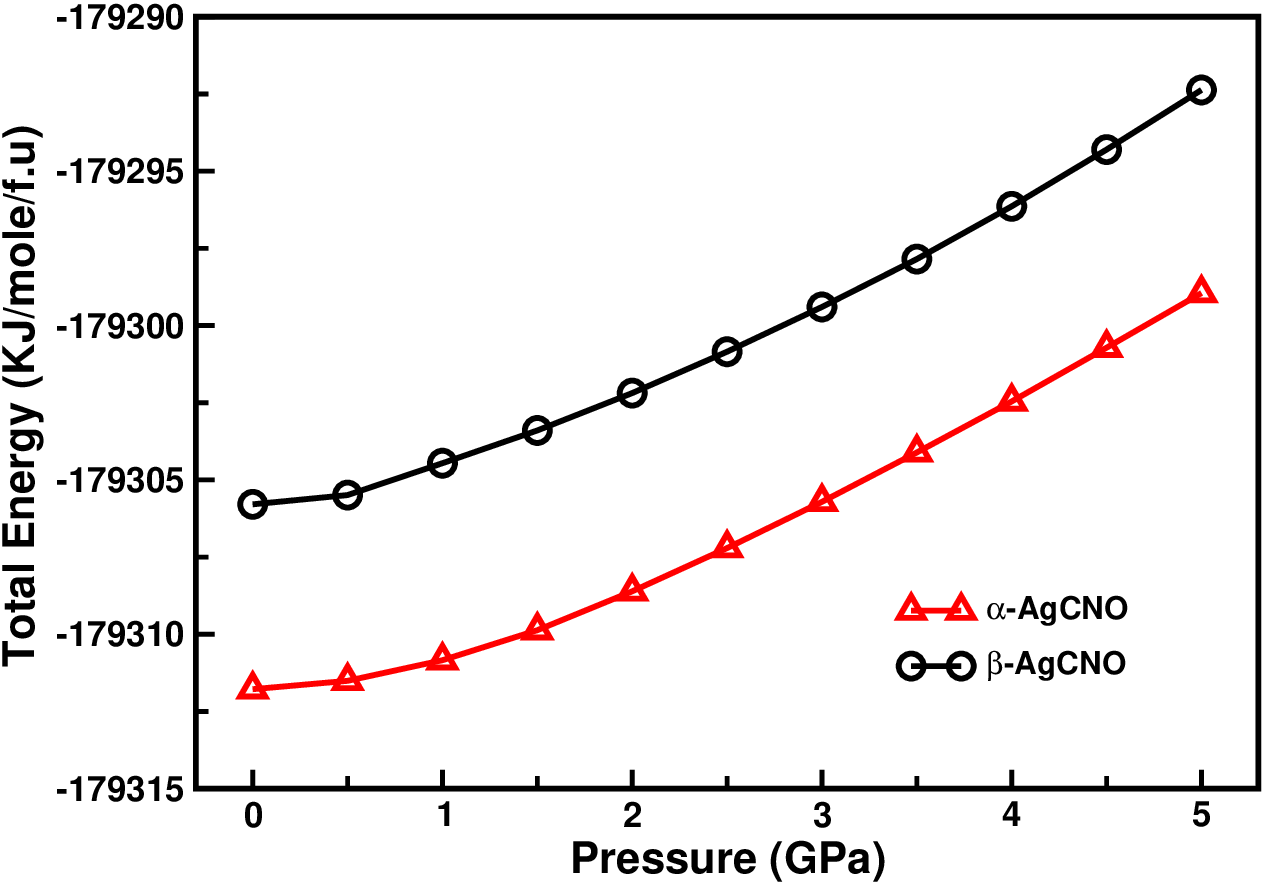}
\caption{(Color online) Calculated total energy as a function of pressure for $\alpha$- and $\beta$- polymorphic phases of AgCNO within DFT-D2 method.} 
\label{PTE}
\end{figure*}

\begin{figure*}
\centering
\includegraphics[height = 3in, width=4in]{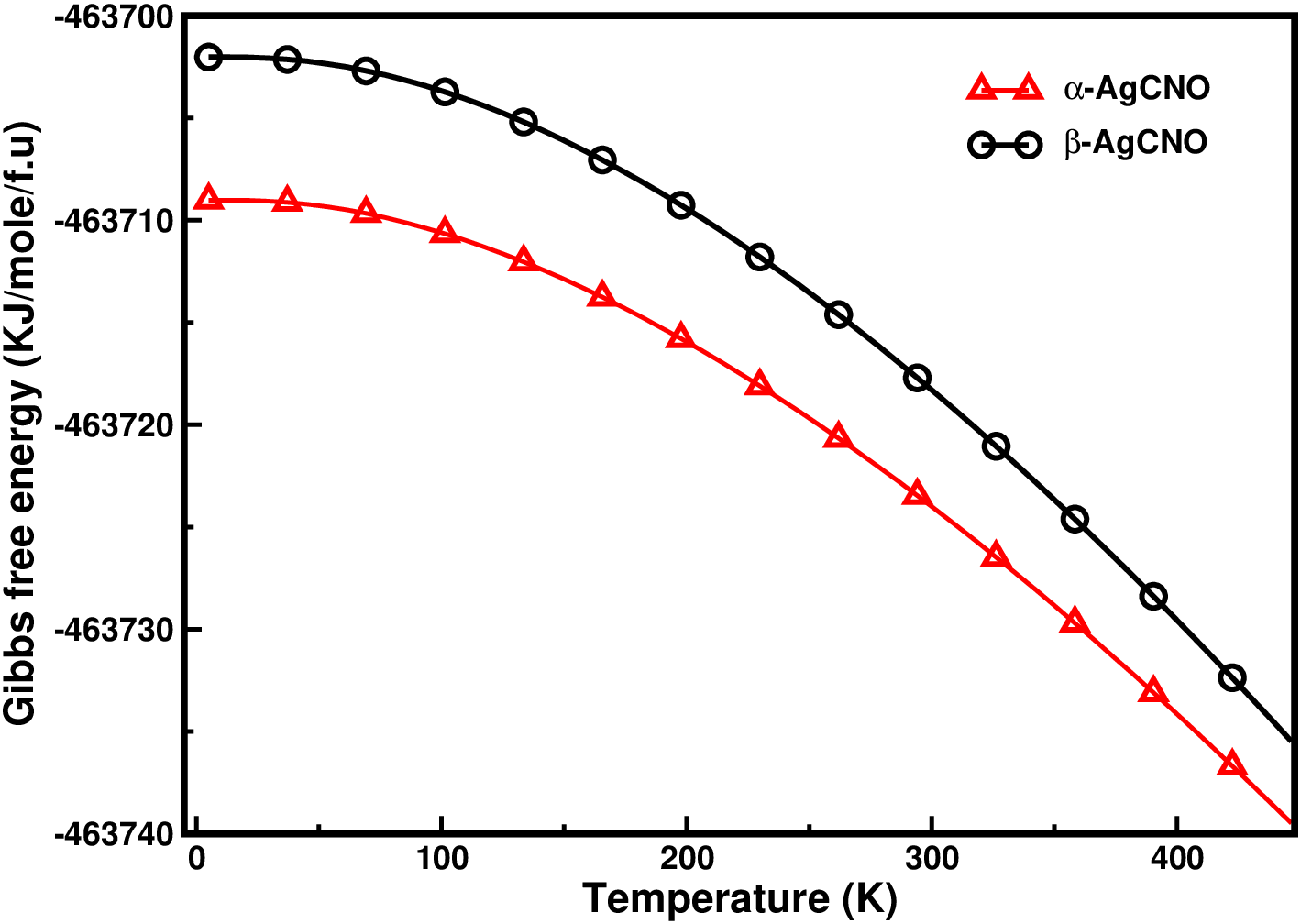}
\caption{(Color online) Calculated Gibbs free energy as a function of temperature for $\alpha$- and $\beta$- polymorphic phases of AgCNO within DFT-D2 method.} 
\label{GT}
\end{figure*}

\begin{figure*}
\centering
\includegraphics[height = 3.0in, width=4.0in]{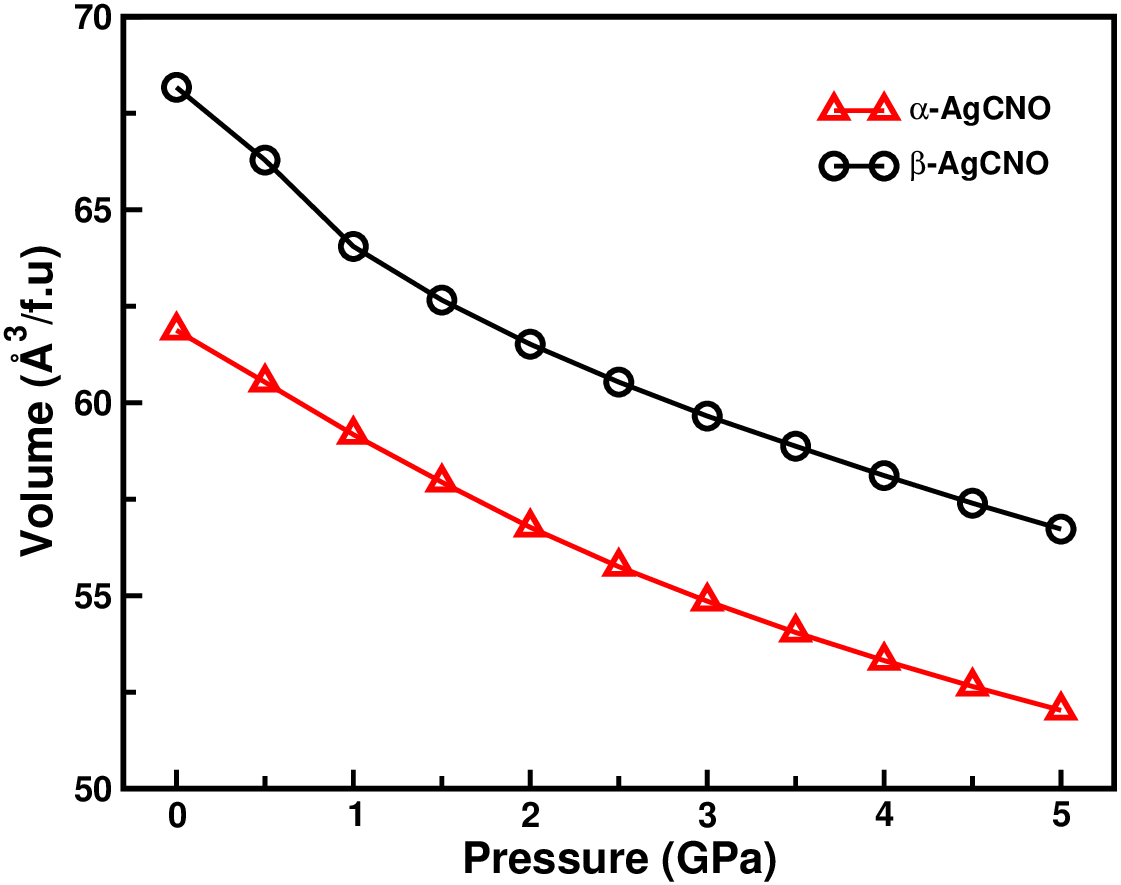}
\caption{(Color online) Calculated volume as a function of pressure for $\alpha$- and $\beta$- polymorphic phases of AgCNO within DFT-D2 method.} 
\label{PV}
\end{figure*}

\begin{figure*}
\centering
\includegraphics[height = 3.0in, width=6.0in]{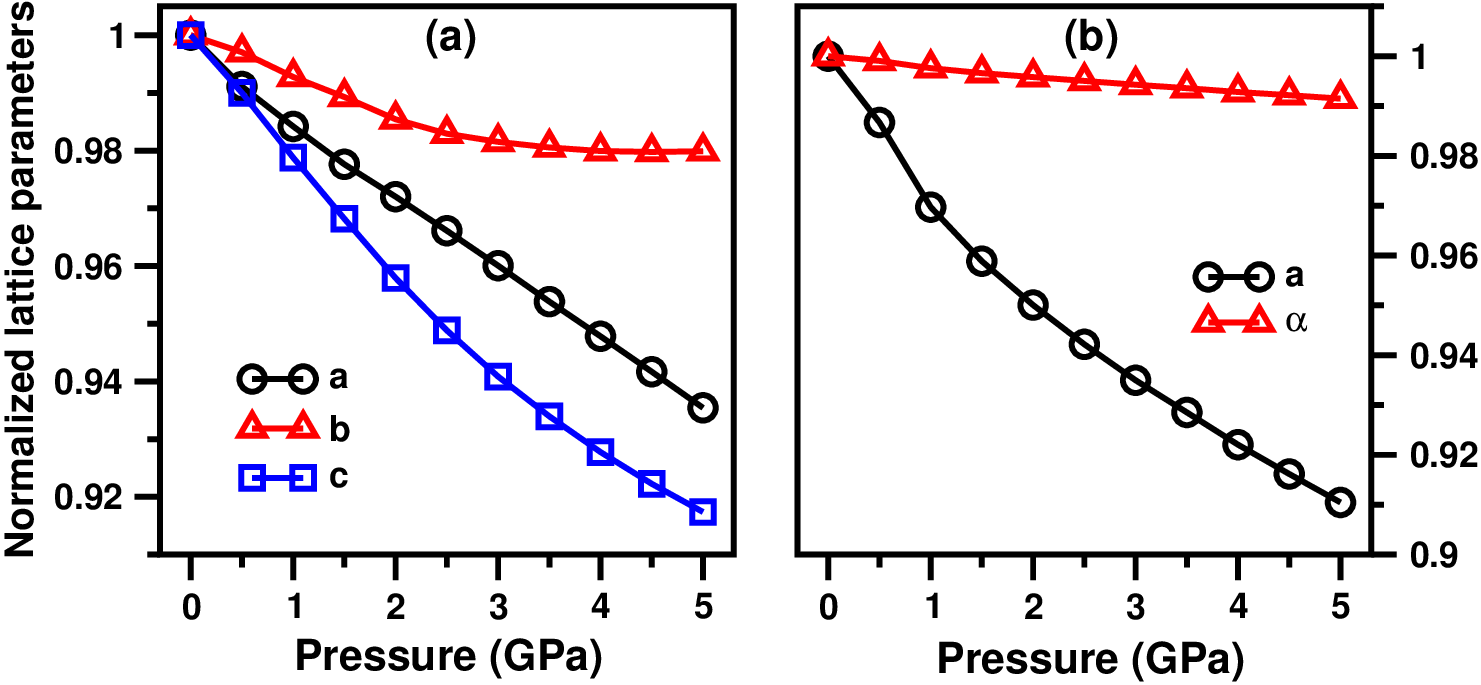}
\caption{(Color online) Calculated normalized lattice parameters of a) $\alpha$- and b) $\beta$- polymorphic phases of AgCNO as a function of pressure within DFT-D2 method.}
\label{abc}
\end{figure*}

\begin{figure*}
\centering
\includegraphics[height = 3.0in, width=6.0in]{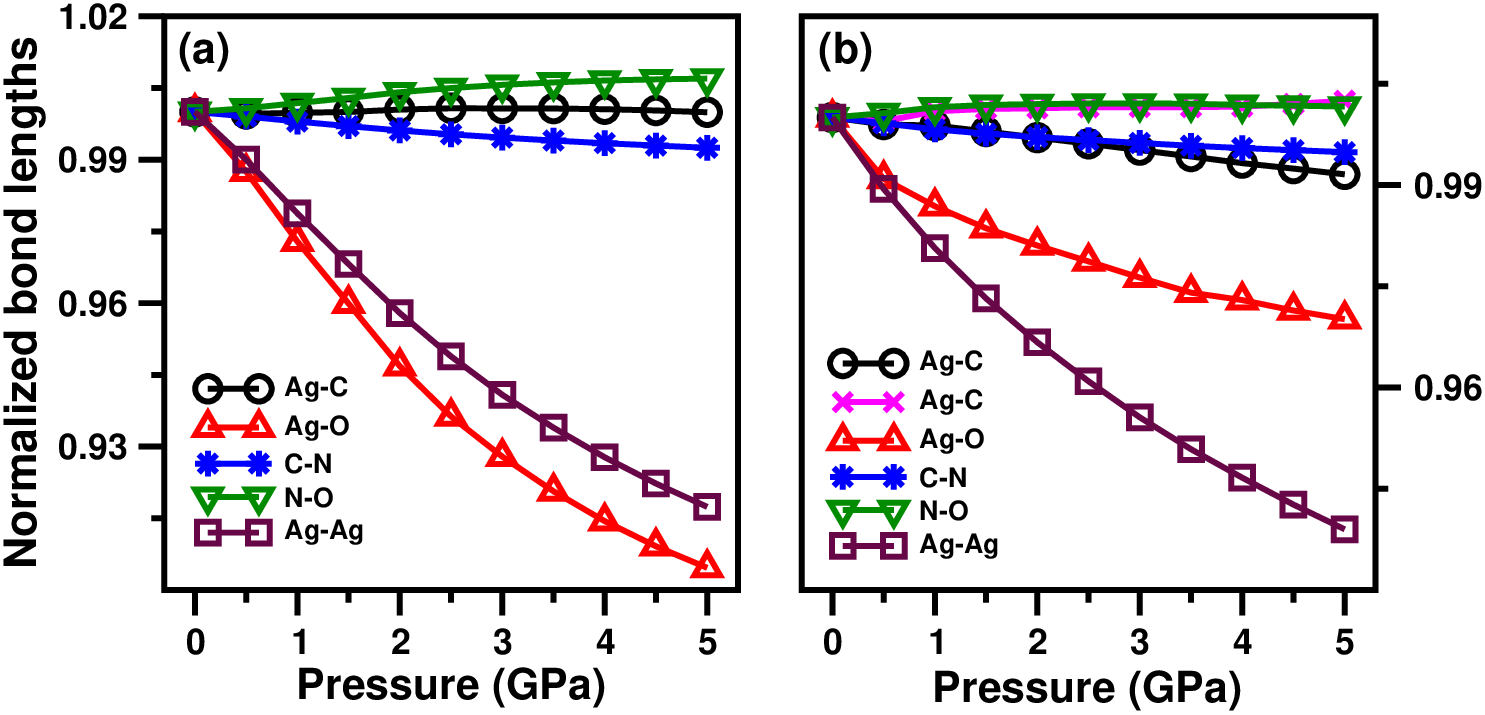}
\caption{(Color online) Calculated normalized bond lengths of  a) $\alpha$- and b) $\beta$- polymorphic phases of AgCNO as a function of pressure within DFT-D2 method.}
\label{BL}
\end{figure*}

\begin{figure*}
\centering
\includegraphics[height = 3.0in, width=6.0in]{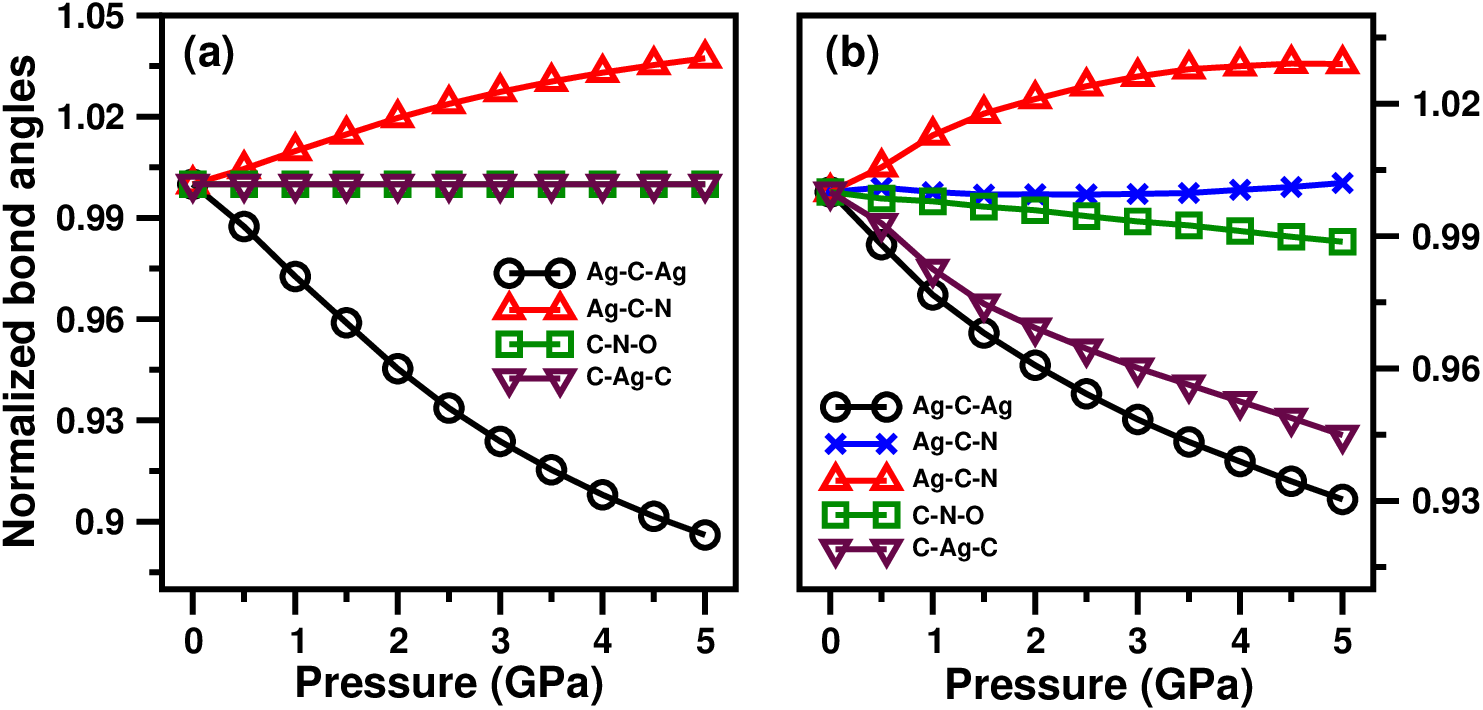}
\caption{(Color online) Calculated normalized bond angles of a) $\alpha$- and b) $\beta$- polymorphic phases of AgCNO as a function of pressure within DFT-D2 method.}
\label{BA}
\end{figure*}

\begin{figure*}
\centering
\includegraphics[height = 3.5in, width=3.0in]{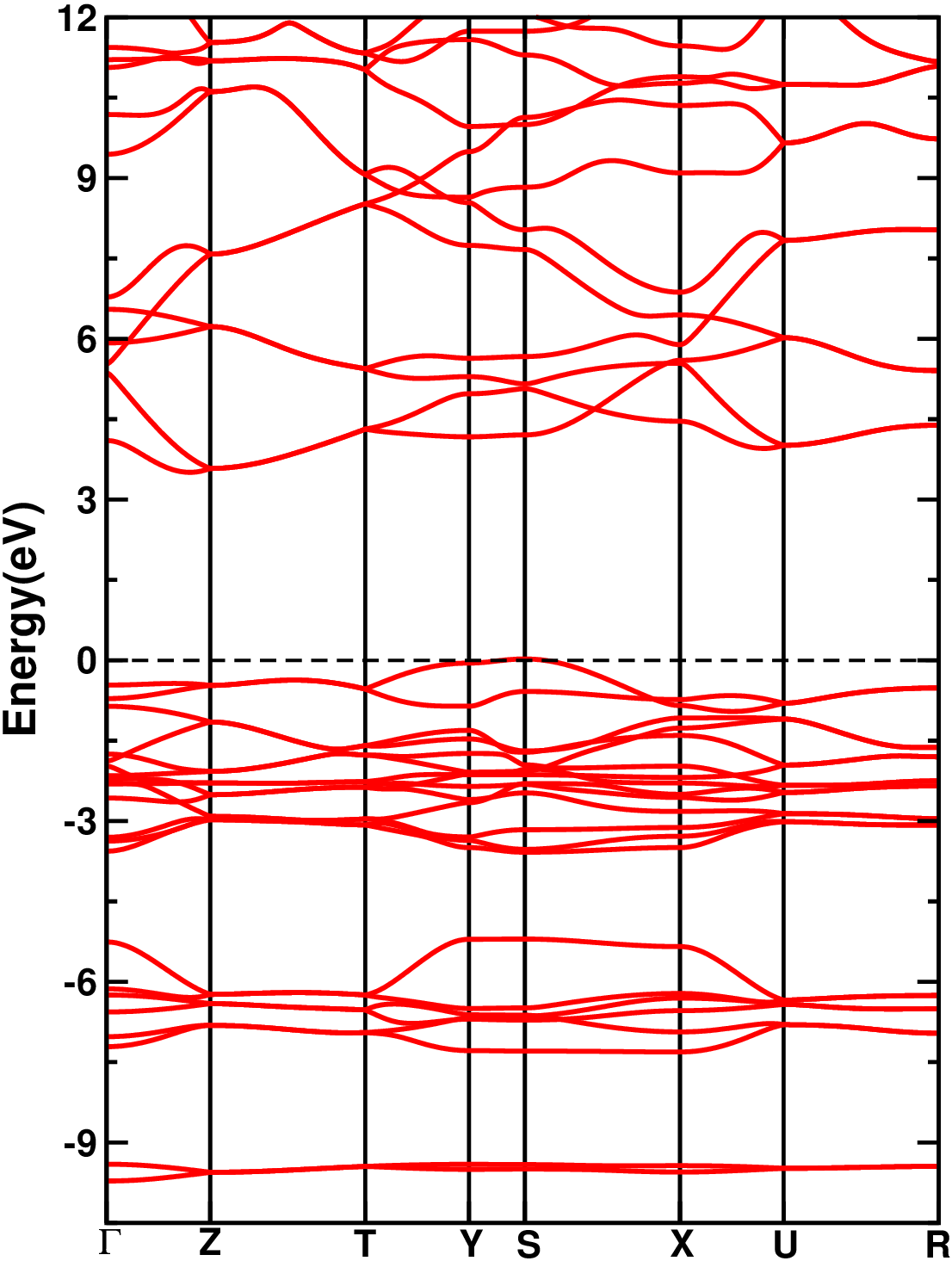} \hspace{0.3 in}
\includegraphics[height = 3.5in, width=3.0in]{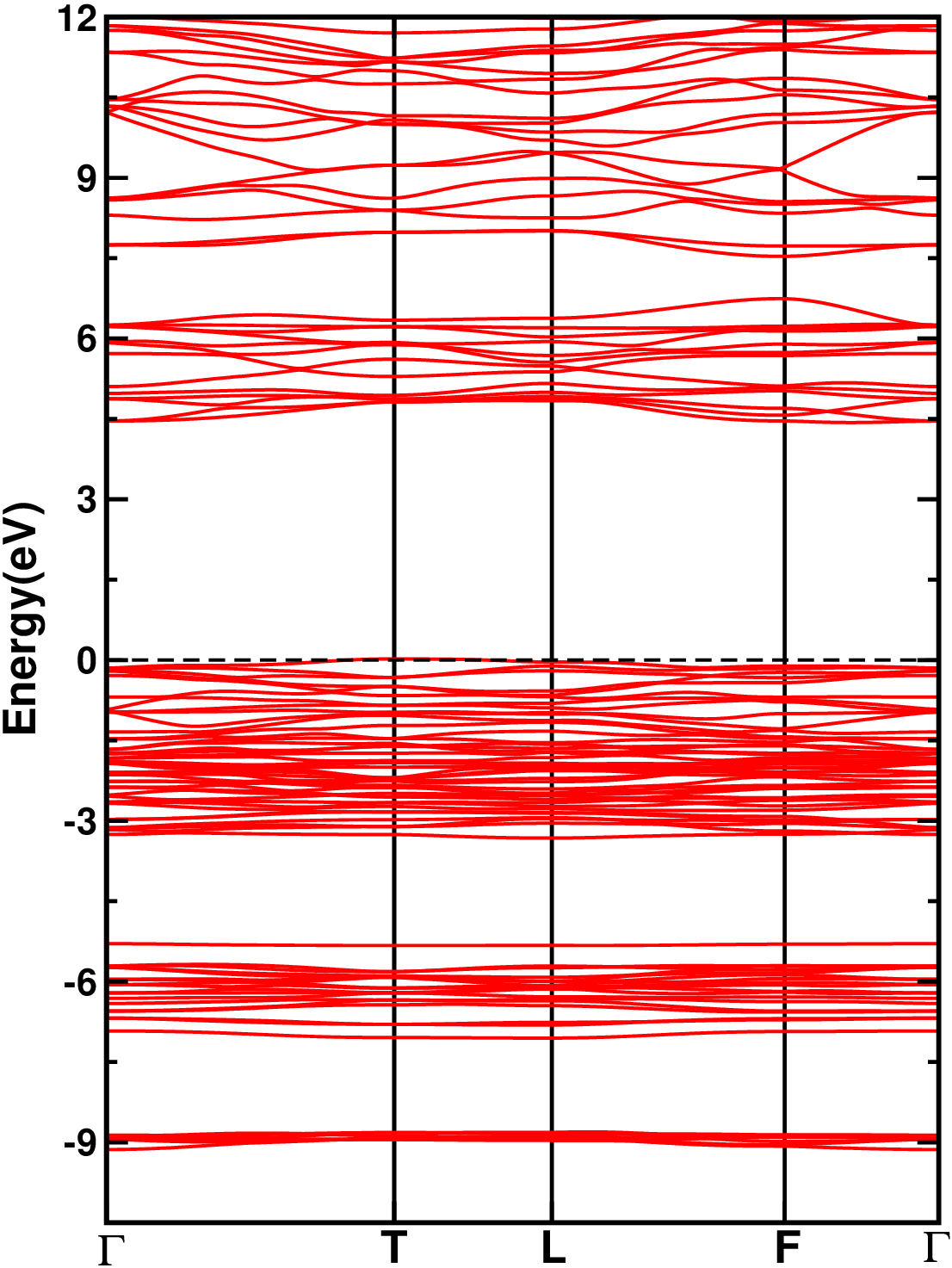}
\caption{(Color online) Calculated electronic band structures of $\alpha$- (left) and $\beta$- (right) phases of AgCNO using TB-mBJ functional at the experimental crystal structures.\cite{barrick,britton2}}
\label{BS}
\end{figure*}

\begin{figure*}
\centering
\includegraphics[height = 3.5in, width=3.0in]{10a.eps} \hspace{0.3 in}
\includegraphics[height = 3.5in, width=3.0in]{10b.eps} 
\caption{(Color online) Calculated total and PDOS of $\alpha$- (left) and $\beta$- (right) phases of AgCNO using TB-mBJ functional at the experimental crystal structures.\cite{barrick,britton2}}
\label{DOS}
\end{figure*}

\begin{figure*}
\centering
\includegraphics[height = 6in, width=6.0in]{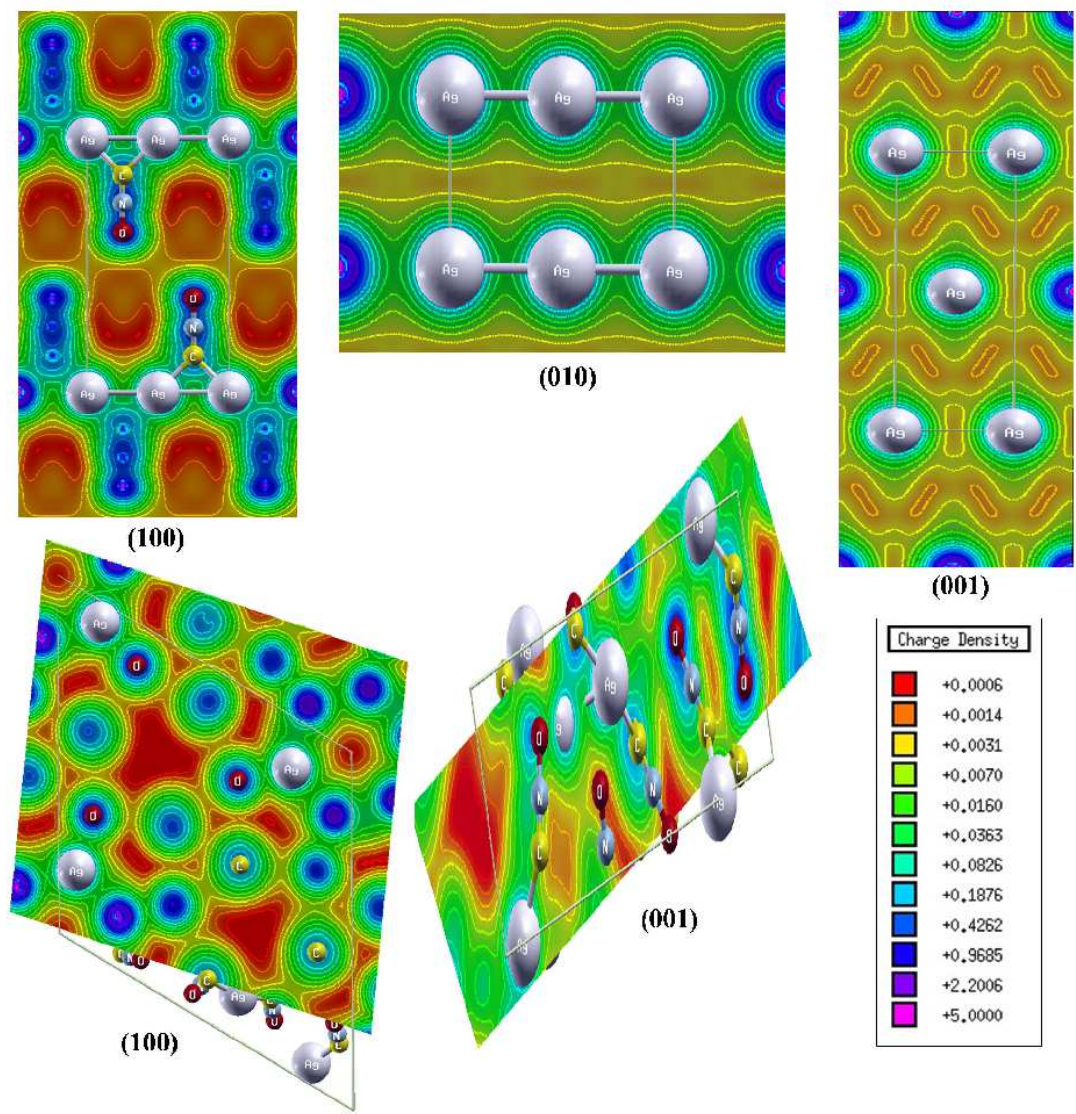}
\caption{(Color online) Calculated electronic charge densities for $\alpha$-phase (top) along (100), (010), (001) and for $\beta$- phase (bottom) along (100), (001) planes.}
\label{charge}
\end{figure*}

\begin{figure*}
\centering
\includegraphics[height = 4.5in, width=6.0in]{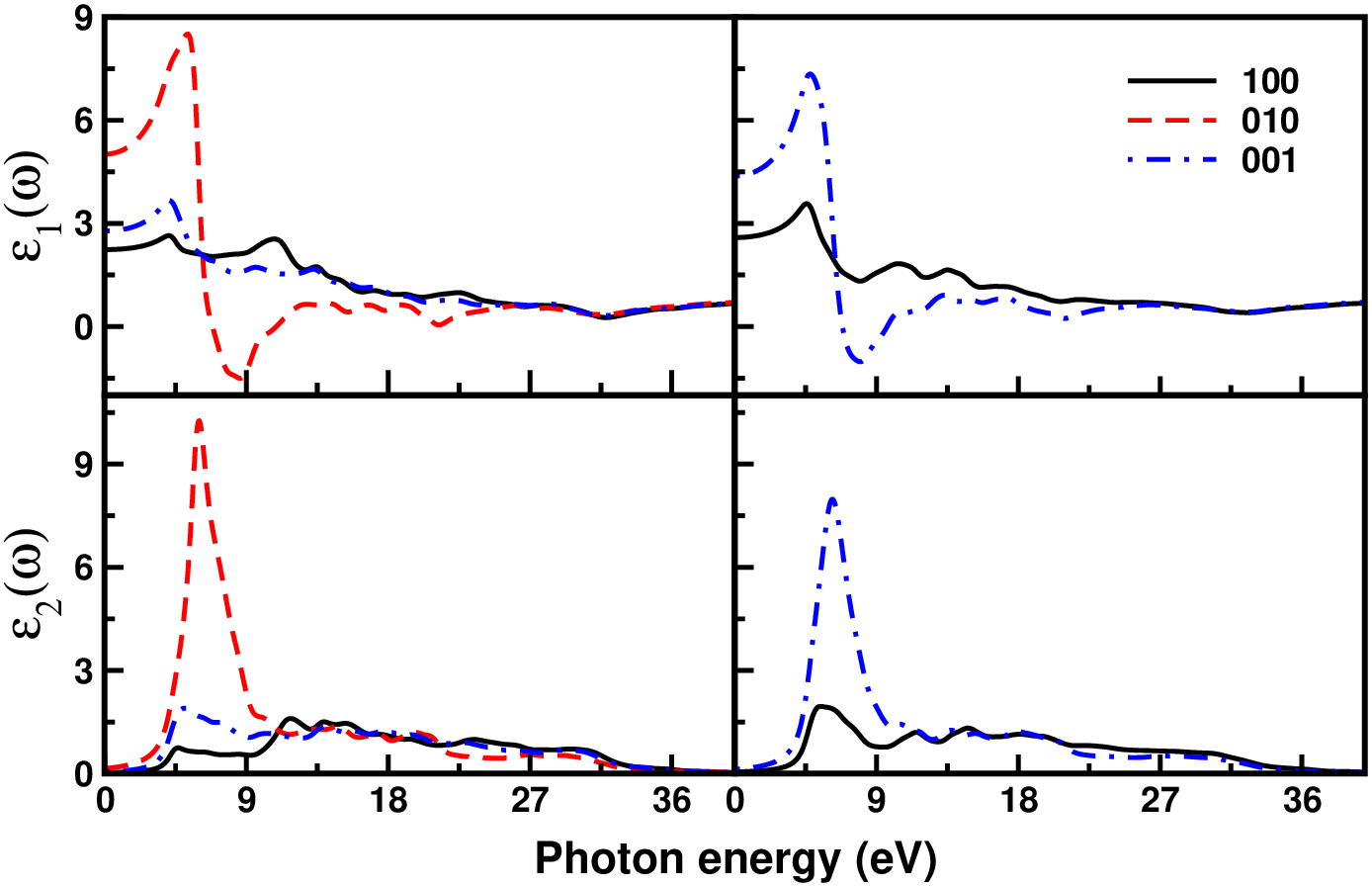}  
\caption{(Color online) Calculated real (top) and imaginary (bottom) parts of dielectric function of $\alpha$- (left) and $\beta$- (right) phases of AgCNO using TB-mBJ functional at the experimental crystal structures.\cite{barrick,britton2}}
\label{epsilon}
\end{figure*}

\begin{figure*}
\centering
\includegraphics[height = 4.5in, width=6.0in]{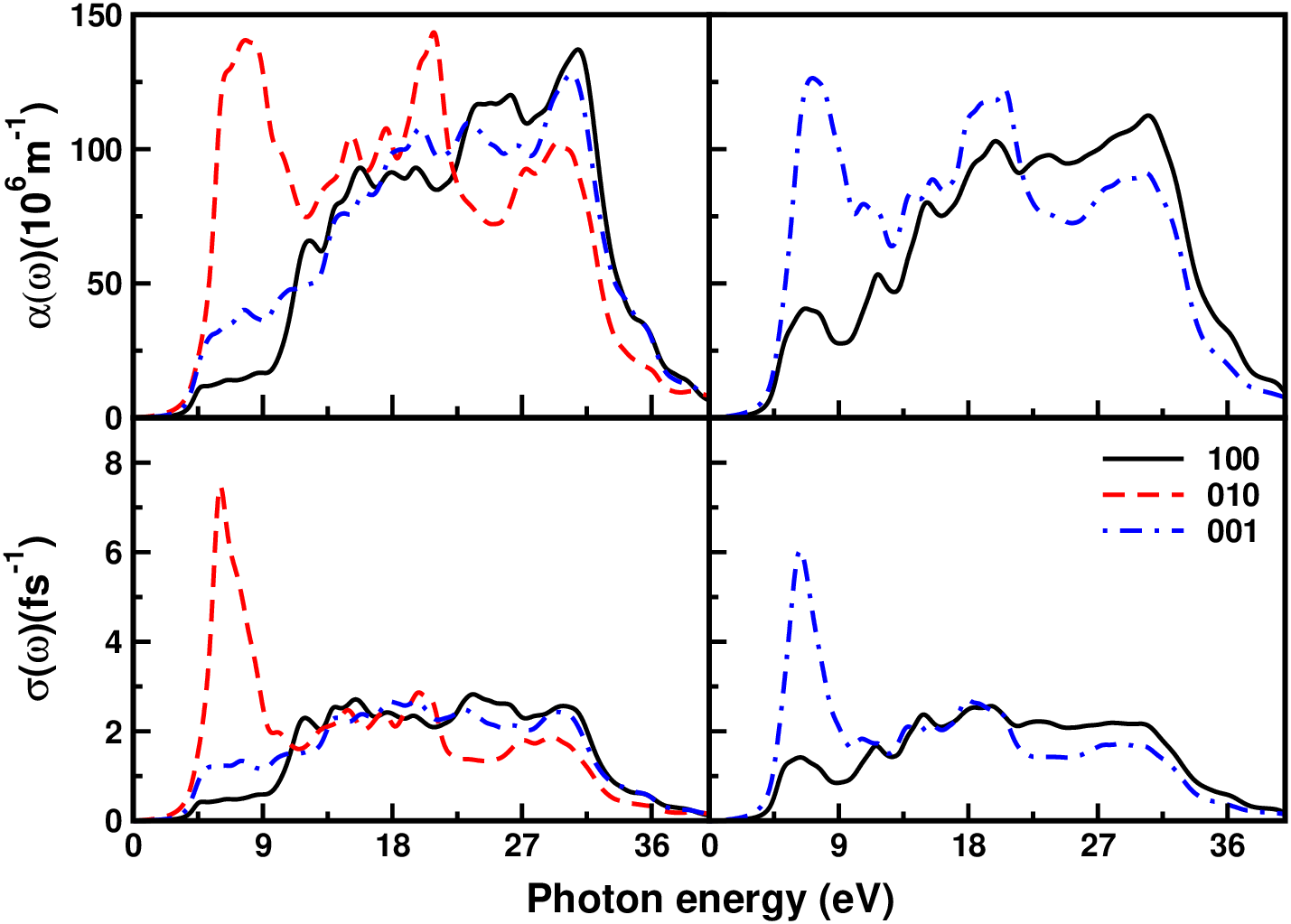}
\caption{(Color online) Calculated absorption (top) and photoconductivity (bottom) of $\alpha$- (left) and $\beta$- (right) phases of AgCNO using TB-mBJ functional at the experimental crystal structures.\cite{barrick,britton2}}
\label{absorp}
\end{figure*}

\end{document}